# Investigating Instabilities in Magnetized Low-pressure Capacitively-Coupled RF Plasma using Particle-in-Cell (PIC) Simulations

Sathya Ganta*[1], Kallol Bera[1], Shahid Rauf[1], Igor Kaganovich[2], Alexander Khrabrov[2], Andrew T Powis[2], Dmytro Sydorenko[3], Liang Xu[4]

[1] Applied Materials, Inc., USA
[2] Princeton Plasma Physics Laboratory, USA
[3] Department of Physics, University of Alberta, AB, Canada
[4] Department of Physical Science & Technology, Soochow University, Suzhou 215006, China
*Corresponding author, email: sathya_ganta@amat.com

## Abstract

The effect of a uniform magnetic field on particle transport in low-pressure radio frequency (RF) capacitively coupled plasma (CCP) has been studied using a particle-in-cell (PIC) model. Three distinct regimes of plasma behavior can be identified as a function of the magnetic field. In the first regime at low magnetic fields, asymmetric plasma profiles are observed within the CCP chamber due to the effect of $\vec{E} \times \vec{B}$ drift. As the magnetic field increases, instabilities develop and form self-organized spoke-shaped structures that are distinctly seen within the bulk plasma closer to the sheath. In this second regime, the spoke-shaped coherent structures rotate inside the plasma chamber in the $-\vec{E} \times \vec{B}$ direction, where $\vec{E}$ and $\vec{B}$ are the DC electric and magnetic field vectors, respectively, and the DC electric field exists in the sheath and pre-sheath regions. The spoke rotation frequency is in the MHz range. As the magnetic field strength increases further, the rotating coherent spokes continue to exist near the sheath. The coherent structures are, however, accompanied by new small-scale incoherent structures originating and moving within the bulk plasma region away from the sheath. This is the third regime of plasma behavior. The threshold values of the magnetic field between these regimes were found not to vary with changing plasma reactor geometry (e.g., area ratio between ground and powered electrodes) or the use of an external capacitor between the RF-powered electrode and the RF source. The threshold values of the magnetic field between these regimes shift toward higher values with increasing gas pressure. This paper provides guidance on the upper limit of the magnetic field for instability-free operation in low-pressure CCP-based semiconductor deposition and etch systems that use the external magnetic field for plasma uniformity control.

# 1. Introduction

Low-pressure radio-frequency (RF) capacitively coupled plasmas (CCP) are used for many applications in the semiconductor industry including etching and deposition. A magnetic field is often used in low to moderate pressure RF CCPs to enhance the plasma density, modify the plasma spatial uniformity, or control plasma chemistry. [1, 2] The influence of the magnetic field on plasma characteristics is primarily due to the magnetization of electrons perpendicular to the magnetic field. The ions are generally unmagnetized in partially ionized plasmas for moderate values of the applied magnetic field. As the gas pressure decreases to a few mTorr, instabilities and rotating structures have been observed in RF CCPs. [3] It is important to avoid such instabilities in processing applications to obtain consistent on-substrate results. The goal of the present study is to identify the operating condition space for stable plasma operation using self-consistent particle-in-cell (PIC) simulations of RF CCPs. Kinetic simulations are essential as fluid and hybrid models for RF CCPs can't capture the kinetic behavior of the instabilities and their impact on charged species transport.

Low-pressure RF CCP magnetized plasmas have been studied using fluid and hybrid plasma models in many previous studies. In Ref. [4], Lieberman et al. determined the RF power transferred to the electrons by the oscillating electron sheath in the presence of a magnetic field using a simple model of a magnetically enhanced reactive ion etching (MERIE) reactor. For a dual-frequency magnetized CCP, Rauf [5] showed that the application of a radial magnetic field enhances charged and neutral species densities in the plasma reactor and the plasma behavior becomes less nonlinear. Using a computational investigation of MERIE sources, Kushner [6] demonstrated that the reduction in electron cross-field mobility decreases the dc bias, which decreases the ion energy and increases ion angular spread at the substrate. For an Ar/ perfluorocyclobutane c-$C_4F_8$/$O_2$ plasma, this trend affects heavier ions more acutely than the lighter ions. The reduction in transverse electron mobility due to the magnetic field can lead to a reversed sheath electric field in both the high and low-frequency discharges in a dual-frequency plasma [7]. For the characterization of three-dimensional magnetized CCPs, Rauf et al. [8] found that the $\vec{E} \times \vec{B}$ drift modifies the plasma structure near the sheaths. Applying an inhomogeneous magnetic field in a very high frequency CCP, Bera et al. [9] demonstrated that both the radial and axial magnetic field contribute to plasma uniformity control at the substrate.

Several PIC models have been reported on magnetized RF CCPs. In [10], Minea et al. showed results of a self-consistent 2D time-resolved PIC model of an RF magnetron discharge which included instantaneous 2D density distributions of electrons and ions. In [11], Turner et al. showed that a small magnetic field (~10G) applied transverse to the electric field will induce a heating mode transition from a pressure-heating dominated state to an Ohmic-heating dominated state. In [12], Zheng et al. demonstrated using a PIC simulation and moment analysis of the Boltzmann equation that Ohmic heating in a low-pressure capacitively coupled magnetized RF discharge is enhanced by the Hall current in the $\vec{E} \times \vec{B}$ direction. With an increasing magnetic field, Ohmic heating dominates the total electron power absorption. Patil et al. [13] and Sharma et al. [14] described the electron bounce cyclotron resonance in magnetized CCPs, which enhances bulk plasma density for a given RF voltage. In [15], Eremin et al. studied the dynamics of plasma in an asymmetric cylindrical RF magnetron with an electric field in the radial direction and a non-uniform magnetic field in the axial direction using PIC simulations. They found that the azimuthal $\vec{E} \times \vec{B}$ drift can accelerate electrons into the inelastic energy range which gives rise to a new mechanism of RF power dissipation related to the Hall current but different from Ohmic heating. Using a 2D PIC-MCC model, Fan et al. [16] studied the plasma behavior in a low-pressure capacitive discharge in the presence of uniform and non-uniform magnetic fields. Their investigation highlighted the role of the magnetic field on the high-energy tail of the electron energy distribution function (EEDF), Ohmic heating, and plasma uniformity. Yang et al. have used 1D PIC models to examine the effect of an inhomogeneous magnetic field on the magnetic asymmetry effect [17] and the secondary electron asymmetry effect [18]. They demonstrated that the magnetic field gradient is an effective method of controlling the self-bias voltage and the ion energy.

Instabilities have been studied in magnetized plasmas, although generally in DC discharges. It is instructive to review instabilities in DC plasmas. Anders et al. [19] measured ion energy distribution functions in a high-power impulse magnetron sputtering (HiPIMS) setup. Their measurements showed broad peaks at several 10s of eV with an extended tail as well as spatial asymmetry due to $\vec{E} \times \vec{B}$ drift. Experimentally, Panjan et al. [20] observed a spoke pattern in the azimuthal direction in the DC magnetron sputtering (DCMS) system with a radial magnetic field and axial electric field. Panjan et al. [3] compared RF magnetron sputtering (RFMS) discharge with the DCMS discharge. In both regimes, stable plasma patterns were observed for a wide range of discharge parameters. For similar gas pressures and discharge powers, the number of spokes in the RFMS regime was always larger than that in the DCMS regime. In Boeuf et al. [21], a 2D PIC simulation in the axial-azimuthal plane was used to describe the development of rotating spoke-like instability in a planar magnetron discharge. The instability was found to be modified Simon-Hoh [22, 23] and it was seen to play an important role in electron heating in the spokes. In Boeuf et al. [24], basic phenomena involving rotating plasma structures in low temperature magnetized plasma were discussed using PIC simulations. Three examples of rotating electron vortices in cylindrical magnetrons, rotating spokes in cylindrical magnetrons, and azimuthal electron-cyclotron drift instabilities in Hall thrusters were described. Xu et al. [25] used a 2D radial-azimuthal PIC simulation of a magnetically enhanced hollow cathode discharge to show the evolution of gradient drift instability into a rotating spoke. The spoke potential hump region was found to be formed because of local anode sheath collapse due to gradient drift instability triggered within the anode sheath.

Instabilities have been studied in simplified models of RF low-temperature plasma sources like Penning discharges. Powis et al. [26] observed rotating plasma spokes in two-dimensional kinetic simulation of a Penning discharge. The similarity between simulation results with and without ionization suggests that the spoke is not the result of an ionization wave. The generated ambipolar electric field and the density gradient are aligned, indicating the presence of collision-less Simon-Hoh instability. In [27], the role of sheath-plasma boundaries on the gradient drift instabilities in partially magnetized plasmas with $\vec{E} \times \vec{B}$ fields was investigated. It was shown that the sheath resistivity may result in long wavelength instabilities which are driven either by the $\vec{E} \times \vec{B}$ drift or the density gradient drifts alone, and in scenarios where the collision-less Simon-Hoh instability is not operative. In [28], a 1D model for cross-field plasma transport in a discharge sustained by uniform RF electric field is derived and validated using a 2D PIC simulation. Lucken et al. [28] show that the effective electron collision frequency depends on the magnetic field if the diamagnetic drift at the sheath edge remains below electron thermal velocity.

In addition to the collision-less modified Simon-Hoh instability, other forms of instabilities are observed in low temperature weakly magnetized plasmas including electron cyclotron drift instability seen in Hall thrusters [20], ion-ion two stream instabilities [29], ion transit time instabilities [30], and modified two stream instabilities [31]. Modified Simon-Hoh instabilities have also been investigated both experimentally and using plasma simulation [32] models before. Kim et al. [33] experimentally investigated the magnetic confinement properties of partially magnetized plasma. The result shows that non-Maxwellian electron energy distribution can enhance anomalous transport by amplifying the electron pressure gradient. With an increase in the magnetic field, the trapping of the energetic electrons in the plasma column strengthens the density gradient and radial electric field, potentially leading to collision-less Simon-Hoh instability. In [34], the generation of a spatially asymmetric sheath structure allowed the beam of energetic electrons to be transported to the extraction region via the $\vec{E} \times \vec{B}$ drift of the electrons. Hara et al. [35] used a fluid dispersion theory in partially magnetized plasmas to examine instabilities where an electric field is imposed perpendicular to a homogeneous magnetic field. They discovered an instability different from the classic modified Simon-Hoh instability under the conditions that the diamagnetic drift is in the direction opposite to the $\vec{E} \times \vec{B}$ drift and the magnitude of the diamagnetic drift is sufficiently larger than the electron thermal speed. Smolyakov et al. [32] theoretically studied instabilities in partially magnetized plasmas using fluid theory. Their nonlinear fluid model captured Simon-Hoh, lower-hybrid, and ion-sound instabilities.

In this study, we consider a 2D capacitively coupled plasma discharge with different electrode sizes in Cartesian geometry. An external magnetic field is applied to the discharge in the direction perpendicular to the planar electric fields. The plasma distribution and instability behaviors are investigated as a function of gas pressure and magnetic field strength. The effect of an external capacitor on the plasma instability is also examined. Also, the effect of RF powered to ground electrode area ratio on plasma instabilities is studied since electrode area ratio affects the size of sheath above RF powered electrode where the instabilities are formed. The computational model is described in Sec. 2. Simulation results are presented in Sec. 3 and instability analysis is given in Sec. 4. Key results are summarized in Sec. 5.

## 2. Computational Model and Setup

EDIPIC, a 2-dimensional (2d3v) electrostatic particle-in-cell (PIC) code has been used for the computational modeling of magnetized RF plasma in this paper. EDIPIC is an open-source code, and it is available with full documentation in public domain on GitHub [36]. EDIPIC employs the explicit Boris scheme in Cartesian geometry for advancing particle position and velocities [37]. The electrostatic Poisson's equation is employed to evaluate the fields with the assistance of the publicly available PETSc library [38]. Monte-Carlo model is included in EDIPIC, and considers elastic, inelastic, and ionization electron-neutral collisions as well as ion-neutral collisions. Collision cross-sections for Argon used in the simulations are from [39]. Simulation of partially ionized plasmas requires crucial atomistic and plasma-surface interaction processes to be included. Surface processes considered in EDIPIC include secondary electron emission from surfaces induced by electrons and ions. Several international benchmarks [40,41] have been previously used to verify EDIPIC.

EDIPIC is written in Fortran 90. Parallelization of the EDIPIC code is done using the Message Passing Interface (MPI) with demonstrated scalability of up to 400 CPU cores. Several diagnostic capabilities are part of the EDIPIC code including the phase space data and ion and electron velocity distribution functions, as well as spectral analysis procedures required to study wave propagation in plasmas which are essential for analyzing RF magnetized plasmas [36].

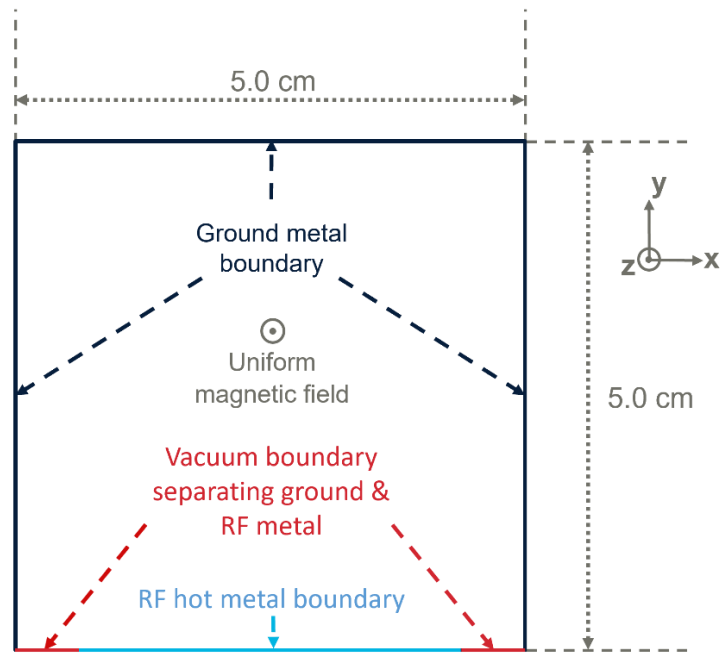

*Figure 1: The 2D Cartesian geometry and boundary conditions used for particle-in-cell (PIC) simulations of RF magnetized plasma.*

Figure 1 shows the geometry setup in the simulations displaying the Cartesian 2D domain with grounded metal boundary walls colored black at the top and sidewalls of the simulation domain. The bottom boundary is an RF-powered electrode metal boundary colored blue at the center and "vacuum gap" boundaries (where voltage is approximated as a linear function inside the gap) colored red on either side of the RF powered electrode to ensure an electrical separation between grounded and powered electrode metal boundaries. The uniform magnetic field is in the +z direction and its magnitude is varied from 0 to 100 Gauss. Neumann boundary condition is used at the vacuum boundary while solving the Poisson equation.

The simulations are done using Argon gas. The dimensions of the Cartesian 2D domain are 5 cm x 5 cm in the x and y directions with no variation assumed in the z-direction. The RF source has a voltage of 200 V (peak amplitude) at 40 MHz. The ion-induced secondary electron emission coefficient $\gamma$ = 0.1 on all metal surfaces and electron induced secondary electron emission is not considered for the RF voltage and pressure range in this work. All particles impinging on the vacuum boundaries are assumed to be lost. The time-averaged electron density ($n_e$), electron temperature ($T_e$), and plasma potential profiles are plotted in the results by time-averaging over an RF cycle where the density/temperature/potential at every spatial point is averaged over 80 regular time intervals within an RF cycle after convergence of the PIC simulation. Convergence took between 100 – 200 μs (depending on the magnetic field) and averaging has been done in the last RF cycle.

The grid-cell size and timestep used in the simulations in this paper are 0.1821 mm and 31.25 ps respectively. These computational parameters ensure that the electron Debye length and plasma frequency are adequately resolved over the range of conditions examined.

## 3. Simulation Results

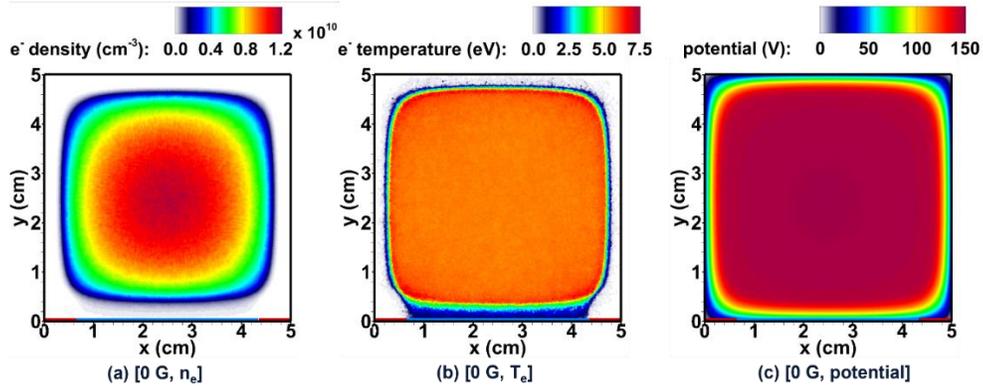

*Figure 2*: The time-averaged (a) $n_e$ (cm$^{-3}$), (b) $T_e$ (eV), and (c) potential (V) profiles obtained from the PIC simulation with Argon gas at 10 mTorr pressure, driven by 200 V amplitude RF voltage at 40 MHz, and with no magnetic field applied. Since there is no capacitor used, no DC self-bias is seen on the RF powered electrode.

The PIC modeling results for the effect of the strength of the external magnetic field, ambient pressure, presence of an external capacitor between the RF-powered electrode and the ground electrode, and RF-powered to ground electrode area ratio are discussed in this section. In these simulations, the pressure is varied from 10 mTorr to 100 mTorr. The width of the RF-powered electrode at the bottom boundary is either the center 75% or 50% of the entire -y boundary width. The widths of the vacuum boundaries remain fixed. This results in having two different RF electrode to ground electrode area ratios: 0.25 and 0.15, respectively. The RF powered electrode at the bottom boundary is either directly connected to the RF source or through a 1 nF capacitor.

We first look at plasma characteristics without an external magnetic field at 10 mTorr pressure. There is no external capacitor in this simulation and the RF powered to ground electrode area ratio is 0.25. The time-averaged electron density ($n_e$) is shown in Figure 2(a). Here, we see plasma density profile has a maximum at the center of the domain forming the bulk plasma region with a few millimeters thick sheath. At low pressure, the plasma diffuses to the center of the process volume creating a center high density profile. The time-averaged electron temperature ($T_e$) and potential spatial profiles are shown in Figures 2(b) and 2(c), respectively. The bulk plasma is at a higher potential compared to the ground electrode or the RF powered electrode which has a 0 time-averaged voltage without an external capacitor. The voltage primarily drops across the sheaths. A DC current of 11.61 mA/m flows from the RF powered electrode to the grounded electrode without the capacitor.

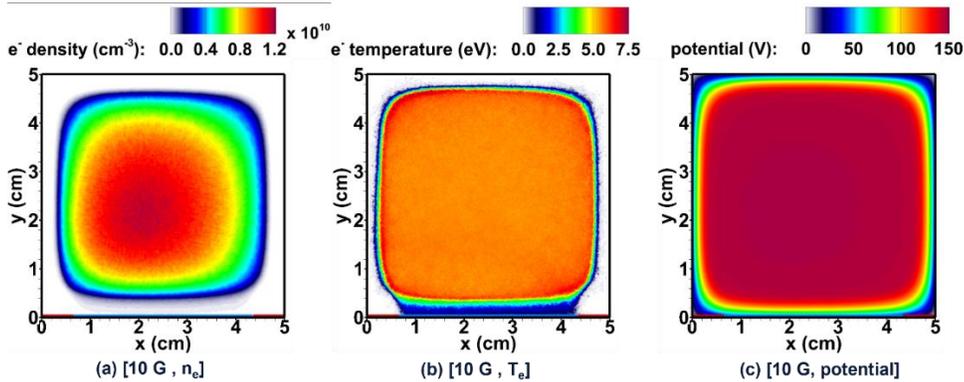

***Figure 3**: The time-averaged (a) $n_e$ (cm$^{-3}$), (b) $T_e$ (eV), and (c) potential (V) profiles obtained from the PIC simulation with Argon gas at 10 mTorr pressure, driven by 200 V amplitude RF voltage at 40 MHz, and a uniform magnetic field of 10 Gauss applied in the +z direction. Since there is no capacitor used, no DC self-bias is seen on the RF powered electrode.*

**3.1 Effect of External Magnetic Field Strength**

The effect of applying a uniform magnetic field of 10 Gauss out of the plane in +z direction on plasma characteristics is shown in Figure 3. The time-averaged electron density profile is shown in Figure 3(a), the time-averaged electron temperature spatial profile is shown in Figure 2(b), and the time-averaged potential profile is shown in Figure 3(c). These results can be compared to the corresponding results in Figure 2 without a magnetic field. Here, we observe a shift of the plasma peak density in the -x direction. This shift is due to the $\vec{E} \times \vec{B}$ drift. The plasma is primarily produced due to stochastically heated electrons at the powered electrode sheath edge and secondary electrons in this electrically asymmetric RF plasma. As these electrons accelerate in the electric field perpendicular to the electrode, they experience the Lorentz force $\vec{F} = q(\vec{v} \times \vec{B})$ in the -x (negative x) direction due to the external magnetic field and hence, move towards the -x direction. The ions experience the same Lorentz force, but their movement is not affected as much as the electrons due to their relatively larger mass. The attractive force between the electrons and ions results in the ions drifting towards the left too, preserving quasineutrality. The electron transport, specifically the diffusion of electrons towards the center of the domain, which is a characteristic of the low-pressure plasma, is hindered by the perpendicular magnetic field causing them to remain closer to the sheath edge where they are produced. This results in an additional -y direction movement of the peak plasma density as the magnetic field strength increases.

The effect of increasing magnetic field on plasma density at 10 mTorr pressure is shown in Figure 4. These simulations are without an external capacitor and the RF powered to ground electrode area ratio is 0.25. The time-averaged electron densities are shown for specific values of magnetic field, 0, 10, 20, 30, 40, 50, 75, and 100 G in Figures 4(a) to 4(h). Three distinct regimes can be identified based on the plasma profiles seen.

**In the first regime** (up to 50 Gauss), we observe a shift towards left bottom corner of the plasma density profile. This movement corresponds to the $\vec{E} \times \vec{B}$ drift experienced by the electrons and ions near the bottom RF powered electrode, as discussed previously. The stronger the magnetic field, the larger the shift. Also, the stronger the magnetic field is, the lesser the transport of electrons by diffusion from the region of production to the center of the domain, causing the peak density to slowly shift in the -y direction. It is observed that the electron density drops with increasing magnetic field strength because a larger drift in the -x and -y directions results in higher plasma loss. The time-averaged (averaged over one RF cycle after convergence) electron current to the left ground wall is found to monotonically increase from nearly 0 at 10 G magnetic field to 4.37 mA/m at 50 G magnetic field, indicating enhanced electron loss. Time-averaged electron current density at the left ground wall at a height of y = 0.5 mm is found to increase from nearly 0 at 10 G to 2.75 mA/m2 at 20 G, 35.63 mA/m2 at 30 G, 92.97 mA/m2 at 40 G and finally to 390.63 mA/m2 at 50 G magnetic field.

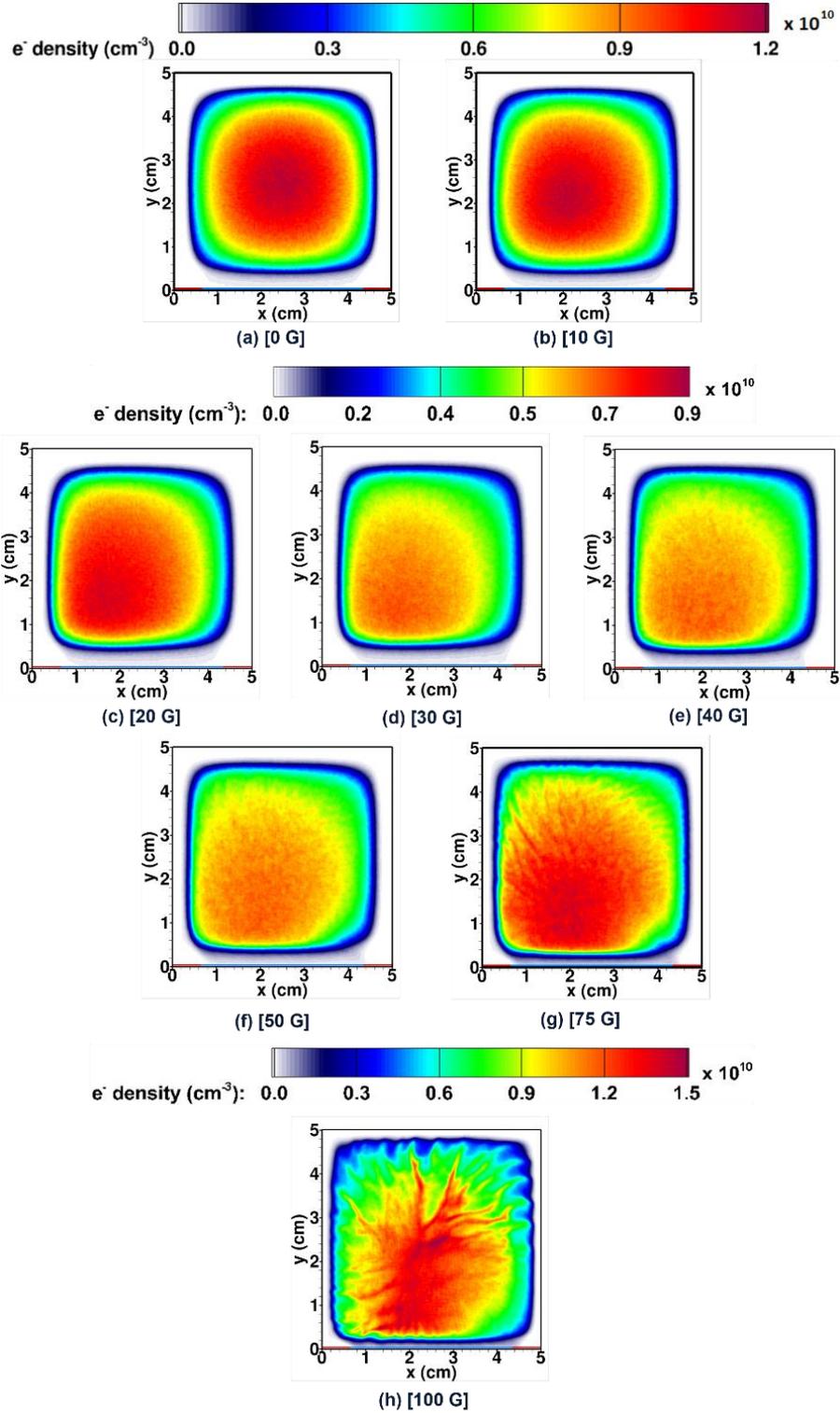

*__Figure 4__: The time-averaged $n_e$ (cm$^{-3}$) profile obtained from the PIC simulation with Argon gas at 10 mTorr pressure, 200 V amplitude RF voltage at 40 MHz, RF powered to ground electrode area ratio of 0.25, $\gamma = 0.1$ from metal boundaries with no external capacitor. A uniform magnetic field in the z-direction is applied with magnitude of (a) 0 Gauss, (b) 10 Gauss, (c) 20 Gauss, (d) 30 Gauss, (e) 40 Gauss, (f) 50 Gauss, (g) 75 Gauss, and (h) 100 Gauss.*

**In the second regime**, at 75 Gauss magnetic field, we observe instabilities in the form of spokes extending from the chamber center to the chamber edges rotating in the anti-clockwise direction. These spokes rotate at frequencies much lower than the RF drive frequency. These structures appear well distinguished everywhere in the sheath and pre-sheath regions except in the bottom region around the RF power electrode boundary. Videos of the rotation of spokes with respect to time are included in the appendix section with a screenshot in Fig. 14. We discuss these spokes in more detail in the next section.

**In the third regime**, at a magnetic field of 100 G, in addition to the rotating spoke-like instabilities within the sheath and pre-sheath regions, we observe instabilities of different sizes moving in an incoherent manner in the center region of the bulk plasma.

The overall electron density increases as magnetic field strength increases beyond 50 G, which is due to reduced electron Larmor radius and better electron confinement by the magnetic field. Drift of the plasma in the -x direction creates a significant new source region for electrons in the sheath edge next to the left ground wall boundary. This can be seen in the ionization rate video for the 75 G case without the capacitor in the appendix, whose time-averaged plot is shown in Figure 19.

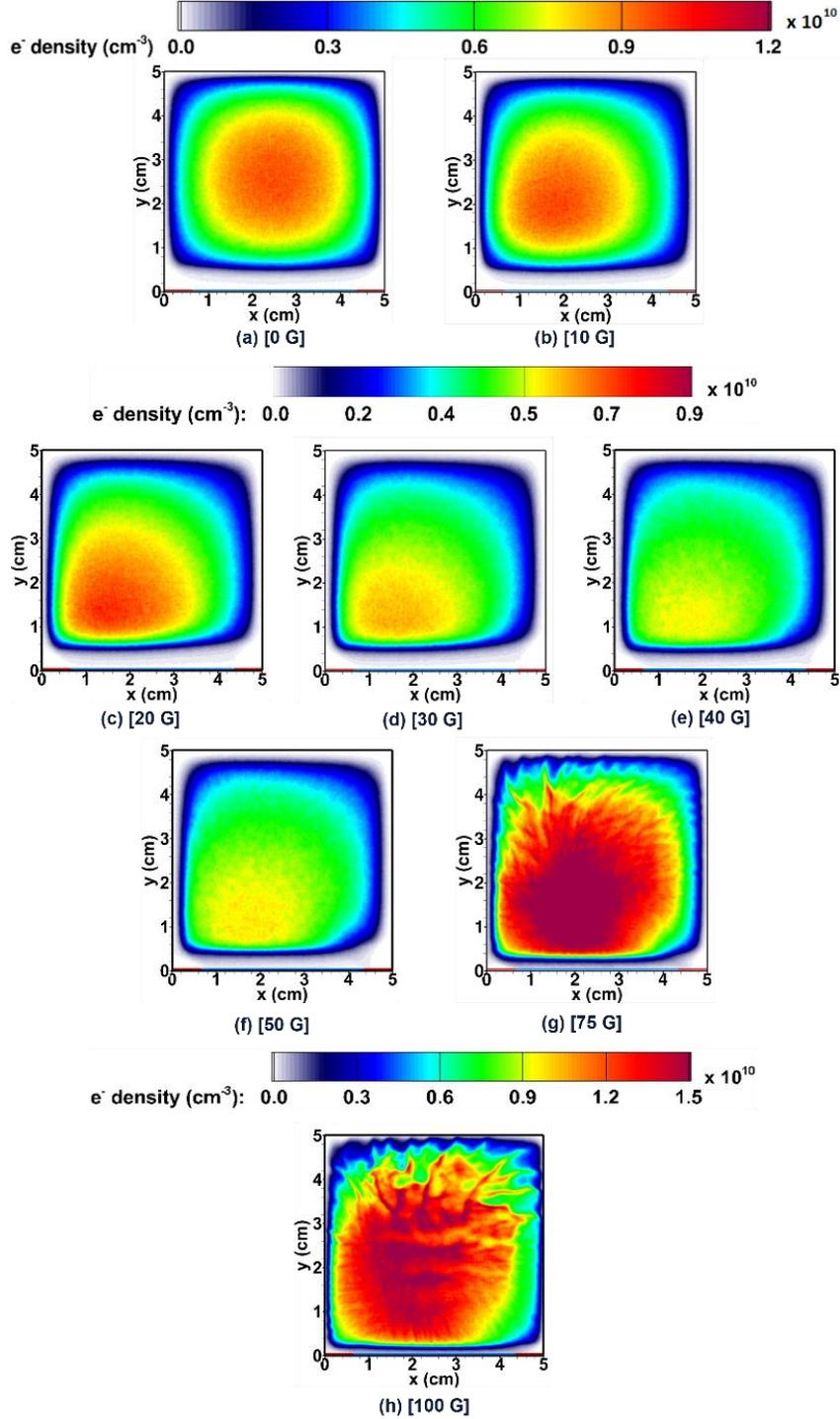

***Figure 5**: The time-averaged $n_e$ (cm$^{-3}$) profile obtained from the PIC simulation with Argon gas at 10 mTorr pressure, 200 V amplitude RF voltage at 40 MHz, RF powered to ground electrode area ratio of 0.25, $\gamma$ = 0.1 from metal boundaries, and 1 nF capacitor connected between RF source and designated RF powered electrode. A uniform magnetic field in the z-direction is applied with a magnitude of (a) 0 Gauss, (b) 10 Gauss, (c) 20 Gauss, (d) 30 Gauss, (e) 40 Gauss, (f) 50 Gauss, (g) 75 Gauss, and (h) 100 Gauss.*

## 3.2 Effect of External Capacitor

Due to unequal areas for the powered and grounded electrodes, a DC current flows through the plasma from RF powered to ground electrodes for the results without a capacitor. One can block this DC current by adding a capacitor in the path of the RF current. The effect of the external capacitor on time-averaged electron density ($n_e$) is shown in Figure 5. These results are for 10 mTorr pressure and an RF powered to ground electrode area ratio of 0.25. The results in Figure 5 can be compared to the corresponding results in Fig. 4. In the first regime below 50 G, the time-averaged electron density is lower when the external capacitor is used and the plasma drifts more in the -x direction. A negative DC voltage builds across the capacitor so that the plasma is electrically more asymmetric and the sheath voltage at the powered electrode is larger with the capacitor. This larger electric field results in stronger $\vec{E} \times \vec{B}$ drift with the capacitor present. The electron transport, specifically the diffusion of electrons towards the center of the domain is hindered by the perpendicular magnetic field causing them to remain closer to the sheath edge where they are formed. This results in an additional -y direction movement of the peak plasma density as magnetic field strength increases like the cases without capacitor. Similar to the cases with no capacitor, the overall electron density decreases with increasing magnetic field strength in the first regime and increases with increasing magnetic field strength in the second and third regimes. Electron density decrease up to 50 G is due to enhanced plasma loss while the subsequent density increase is because of better electron confinement by the magnetic field.

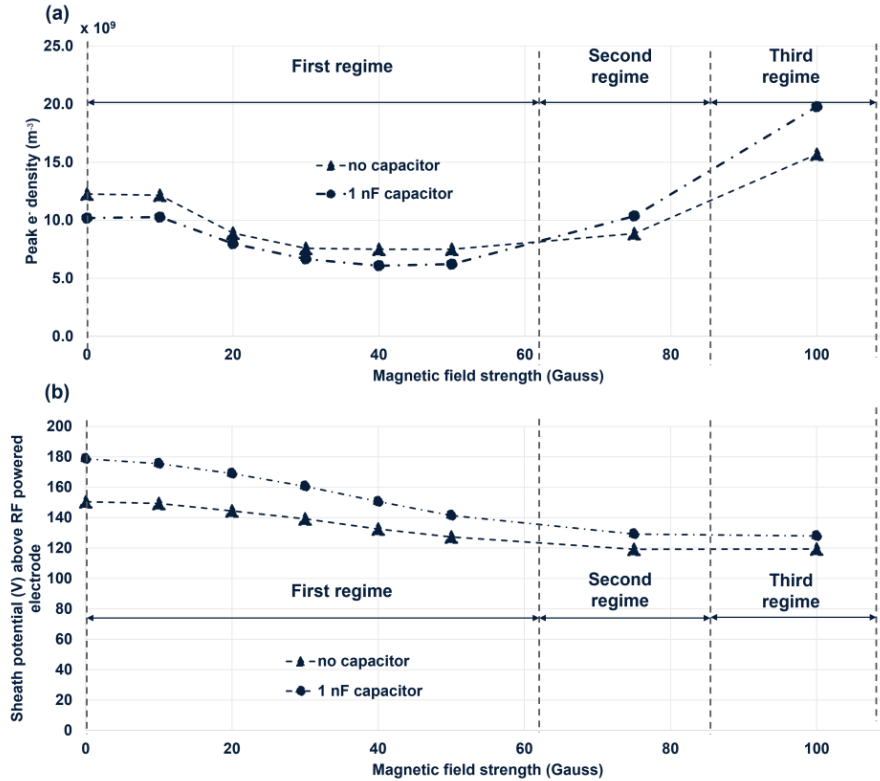

***Figure 6***: *(a) Peak time-averaged $n_e$(cm$^{-3}$) and (b) time-averaged sheath potential (V) above RF powered electrode in the simulation domain obtained from the PIC simulation vs z-directional uniform magnetic field strength for Argon gas at 10 mTorr pressure, 200 V amplitude RF voltage at 40 MHz, RF powered to ground electrode area ratio of 0.25, $\gamma$ = 0.1 from metal boundaries, and with no external capacitor and a 1 nF external capacitor. The magnetic field boundaries between the first, second, and third regimes are shown.*

The variation of peak electron density with increasing magnetic field both with and without an external capacitor is shown in Figure 6(a). The variation of sheath potential above the RF powered electrode with increasing magnetic field strength both with and without an external capacitor is shown in Figure 6(b). In Figure 6(b), we can observe that the sheath potential above the RF powered electrode in the first regime is larger with the addition of a capacitor. The peak electron density is also smaller with the capacitor, as seen in the first regime in Figure 6(a).

The addition of the capacitor does not change the spoke and incoherent structures observed in the second and third regimes. The capacitor also does not change the threshold value of the magnetic field between the different regimes. The 1 nF capacitor cases in second and third regime have higher peak electron density compared to the corresponding cases with no external capacitor. A comparison of ionization rates between the 75 G case with and without the capacitor in Figures 19 and 20 in the appendix shows that the case with a capacitor has a stronger ionization source compared to the case without a capacitor, which is responsible for higher peak electron density with the capacitor in the second and third regimes. Figure 20 in the appendix shows the ionization rate video for the 75 G case with 1 nF capacitor.

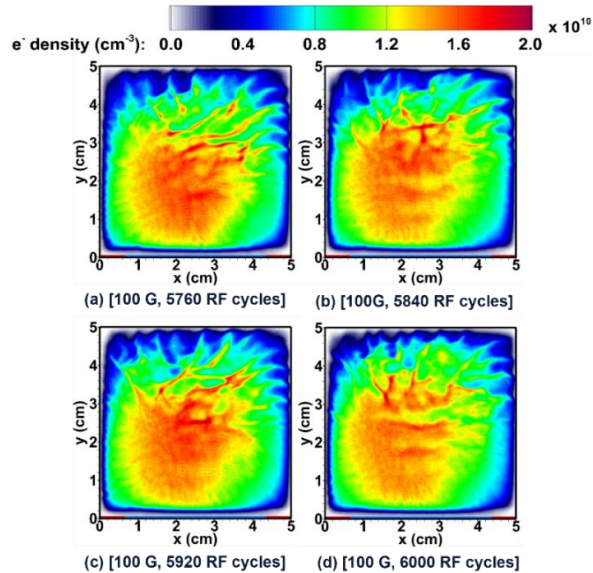

*Figure 7*: *The time-averaged (averaged over 1 RF cycle) $n_e$ (cm$^{-3}$) profile at 4 different times namely (a) 5760 RF cycles; (b) 5840 RF cycles; (c) 5920 RF cycles; and (d) 6000 RF cycles for Argon plasma at 10 mTorr pressure driven by 200 V amplitude RF voltage at 40 MHz with a uniform magnetic field strength of 100 Gauss.*

In Figure 7, the time-averaged electron density profiles corresponding to case with 100 G magnetic field, 10 mTorr pressure, a 1 nF blocking capacitor, and RF powered electrode to ground electrode area ratio of 0.25 are shown. The four Figures 7(a)-(d) show time-averaged (averaged over 1 RF cycle) electron density profiles after the elapse of 5760, 5840, 5920 and 6000 RF cycles (1 RF cycle = 1/(40 MHz) = 25 ns) respectively where unstable structures of different shapes and sizes can be observed moving around in the central bulk plasma region of the chamber in different directions in an incoherent manner. The video for plasma density variation with time showing these instabilities is included in the appendix section with a screenshot in Fig. 16.

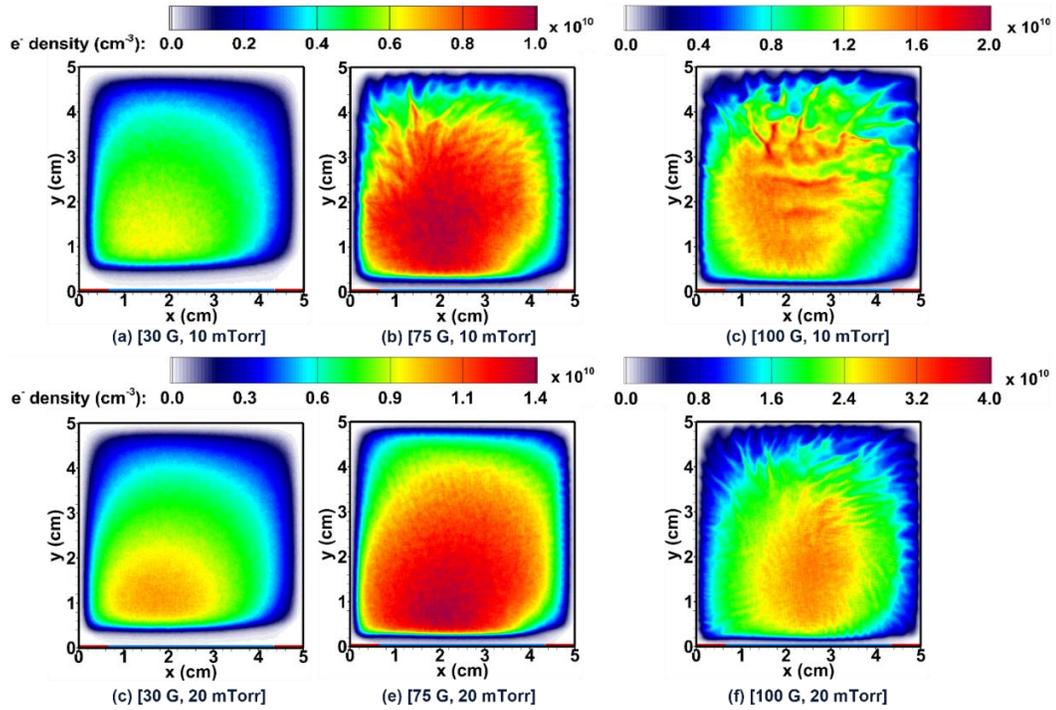

*Figure 8: The effect of gas pressure on time-averaged $n_e$ ($m^{-3}$) profiles comparing 10 mTorr and 20 mTorr pressures at 3 different uniform magnetic field strengths of 30 Gauss, 75 Gauss and 100 Gauss applied in +z-direction for Argon plasma driven by 200 V amplitude RF voltage at 40 MHz.*

**3.3 Effect of Pressure**

We next examine the effect of pressure on plasma characteristics. In Figure 8, $n_e$ is compared at two pressures (10 and 20 mTorr) for magnetic field strengths of 30, 75, and 100 G. These results are for a 1 nF blocking capacitor at the RF powered electrode with an RF powered to ground electrode area ratio of 0.25. We can see that the magnetic field threshold between the first regime of $\vec{E} \times \vec{B}$ drift and the second regime of self-organized instabilities changes from less than 75 G at 10 mTorr pressure to a value between 75 and 100 G at 20 mTorr pressure. Also, there are no incoherent structures within bulk plasma up to 100 G for 20 mTorr pressure indicating that the threshold between the second and third regimes of plasma behavior has moved to a magnetic field value higher than 100 G at 20 mTorr pressure. At 100 Gauss, spokes have started forming at 20 mTorr but they are much weaker than that at 10 mTorr. The video of $n_e$ vs time at 20 mTorr and 100 G is included in the Appendix with a screenshot in Fig. 17. The effect of pressure can be attributed to the effect of electron-neutral collisions, which dampen the instabilities at the higher pressure [28].

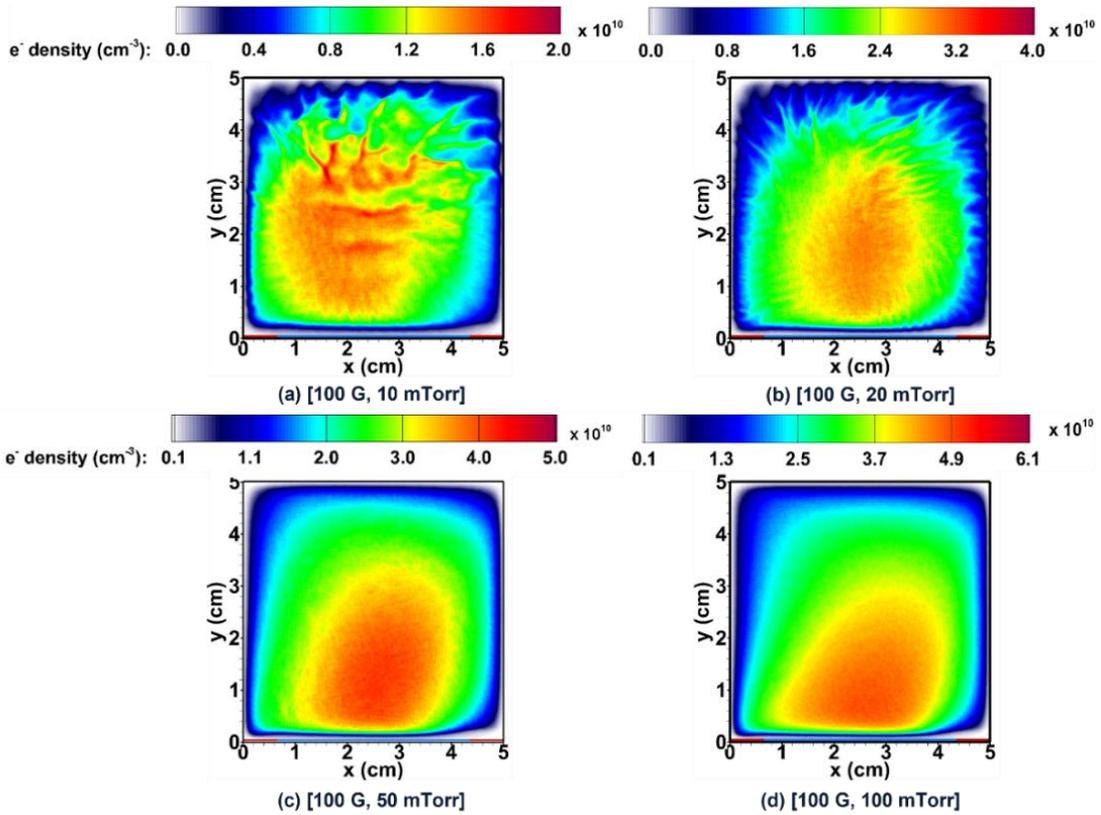

*Figure 9*: The time-averaged $n_e$ (cm$^{-3}$) profiles at 4 different pressures namely (a) 10 mTorr, (b) 20 mTorr, (c) 50 mTorr, and (d) 100 mTorr for Argon plasma driven by 200 V amplitude RF voltage at 40 MHz with a uniform magnetic field of 100 Gauss applied in the +z-direction.

Exploring the effect of pressure further, in Figure 9, the electron density profile, $n_e$ is shown for 100 G comparing four different pressures between 10 and 100 mTorr. These results are for a 1 nF capacitor at the RF powered electrode and an RF powered to ground electrode area ratio of 0.25. We can see that the plasma behavior for the same magnetic field strength changes from incoherent instabilities within the bulk plasma (third regime) at 10 mTorr pressure to structured self-organized instabilities (second regime) at 20 mTorr to the first regime ($\vec{E} \times \vec{B}$ drift) at 50 and 100 mTorr pressures.

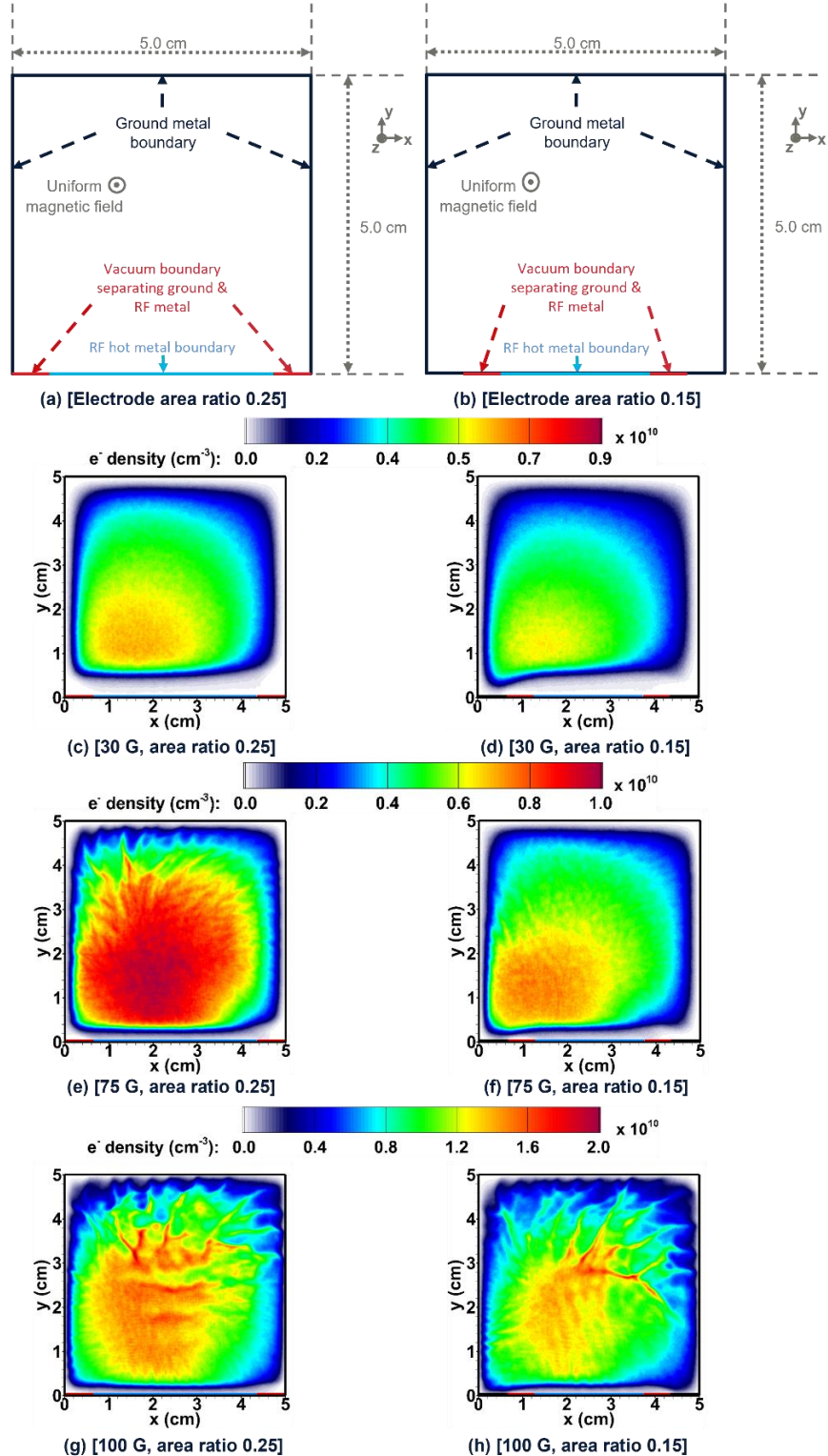

*__Figure 10__: The effect of changing electrode area ratio on time-averaged $n_e$ (cm$^{-3}$) profiles comparing RF powered to ground electrode area ratios of (a) 0.25 and (b) 0.15 respectively at 10mTorr pressure and at 3 different uniform magnetic field strengths of (c, d) 30 Gauss, (e, f) 75 Gauss and (g, h) 100 Gauss applied in +z-direction for Argon plasma driven by 200 V amplitude RF voltage at 40 MHz.*

### 3.4 Effect of RF Powered to Ground Electrode Area Ratio

We next consider the effect of RF powered to ground electrode area ratio. The two geometries corresponding to RF powered to ground electrode area ratios of 0.25 and 0.15 are shown in Figures 10(a) and 10(b), respectively. In Figure 10(c)-(h), $n_e$ is shown at 10 mTorr pressure corresponding to RF powered to ground electrode area ratios of 0.25 and 0.15 at 30, 75, and 100 G for these geometries. These results are for a 1 nF capacitor at the RF powered electrode. The plasma profiles for all magnetic field strengths change both in magnitude and in shape as the RF powered to ground electrode area ratio is decreased. With a smaller RF powered to ground electrode area ratio, the DC bias is larger, and the sheath is thicker above the RF powered electrode. As the left ground wall extends to the bottom -y boundary in cases with the smaller electrode area ratio, the plasma also moves closer to this extended ground electrode near the bottom-left corner as seen in Figure 10. This shift is because the sheath thickness next to the larger area ground wall is expected to be smaller than the sheath thickness above smaller area RF powered electrode. With a thicker sheath above RF powered electrode, bulk plasma volume reduces leading to a smaller volumetric source, which makes the plasma weaker for the same RF voltage. The plasma instability behavior and the threshold values between different regimes are not affected by the RF powered to ground area ratio. The video of $n_e$ vs time for the 75 G case with 1 nF capacitor in the second regime at 10mTorr where the electrode area ratio is 0.15 is included in the Appendix with a screenshot in Fig. 18.

## 4. Analysis of Simulation Results for Instability

In this section, results from the PIC simulations are analyzed to determine the nature of the instability observed at high magnetic field and low gas pressure. We consider the case at 10 mTorr gas pressure, 75 G magnetic field, 1 nF blocking capacitor at the RF powered electrode, and RF powered to ground electrode area ratio of 0.25. For this case, the time-averaged electric field is strongest on top of the RF powered electrode and along the vacuum gaps, and there is a shift of plasma density profile towards the -x boundary. Time-averaged electron density ($n_e$) for these conditions is shown in Figure 5(g) and a video of $n_e$ changing with time is included in the appendix with a screenshot in Figure 15.

In this regime, we notice that the sheath is thin in the bottom left region. The sheath thickens starting from the bottom right corner going all the way around the chamber in the anti-clockwise direction along the boundaries until we reach the thin sheath region at the left bottom. From movies in Fig. 15 we notice that the spoke-shaped unstable structures originate in the bottom right region where there is a thick sheath and move around in the thick sheath region in an anti-clockwise direction until they reach the thin sheath region in the bottom left where the spoke-like structures disappear. The anti-clockwise direction is the direction of diamagnetic drift i.e., $\vec{\nabla} n_e \times \vec{B}$ where $\vec{\nabla} n_e$ is the electron density gradient. The $\vec{E} \times \vec{B}$ drift is in the clockwise direction and is smaller in magnitude compared to the diamagnetic drift.

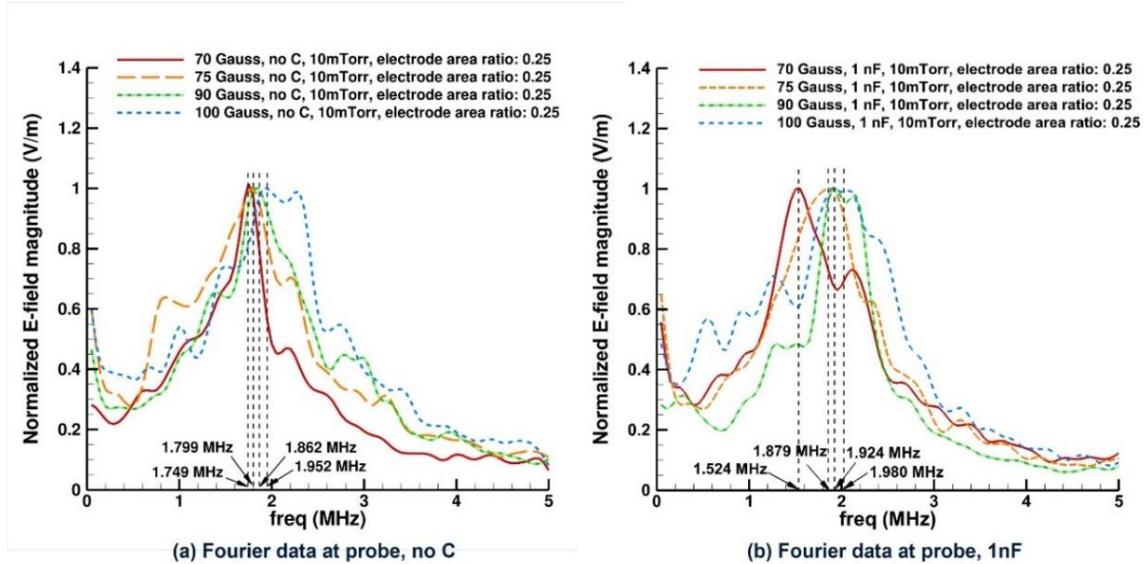

*Figure 11*: The Fourier analysis of time-domain electric field magnitude data collected at the probe position in the top-left corner (x=1.67 cm, y = 3.33 cm) from the PIC simulations for varying magnetic field of 70 G, 75 G, 90 G, and 100 G applied in +z-direction both for (a) no capacitor and (b) 1 nF capacitor for Argon plasma driven by 200 V amplitude RF voltage at 40 MHz.

A probe at x = 1.67 cm, y = 3.33 cm was used to analyze the frequency of movement of waves associated with the instabilities. This probe is chosen as it is in the top left corner of the geometry in the sheath region and in the region where distinct organized rotating instabilities are observed in both second and third regimes. Fourier analysis of time-domain electric field obtained at this probe position for magnetic fields of 70, 75, 90, and 100 Gauss at 10 mTorr pressure is plotted in Figure 11 with and without the 1 nF capacitor. These results show that the waves of coherent spoke-like structures have a continuous spectrum peaking at 1.879 MHz (marked in Figure 11(b)). The peak occurs between 1.5 and 2 MHz for the rest of the cases considered in Figure 11. While the spokes rotate at much lower frequencies compared to the RF frequency, the electron cyclotron frequency at a magnetic field strength of 75 Gauss is 1.32 GHz. This eliminates the possibility of the electron cyclotron drift instabilities (ECDI) which occur at frequencies close to the electron cyclotron frequency [42].

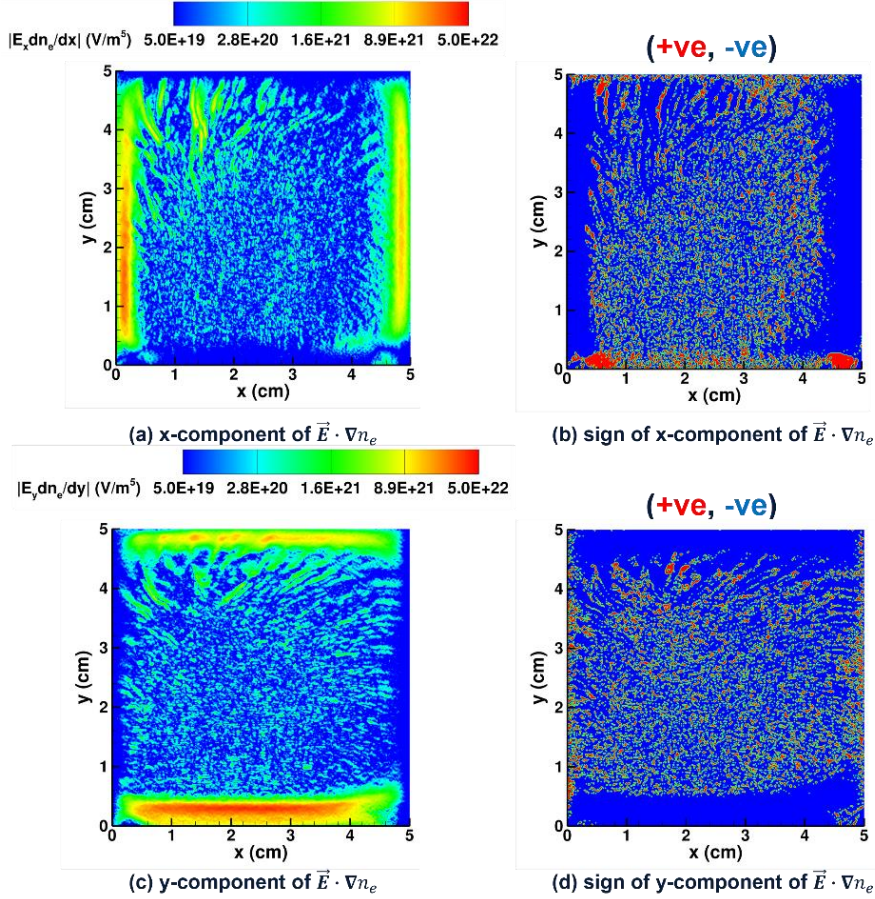

*Figure 12*: *The x and y-components of the dot product of the time-averaged electric field $\vec{E}$ (in V/m) and the time-averaged electron density gradient vector $\nabla n_e$ (in m$^{-4}$) with their magnitudes plotted in (a) and (c) respectively and their signs plotted in (b) and (d) respectively for the 75 Gauss, 10 mTorr, and 1 nF case.*

Another possibility is the modified Simon-Hoh instability [22, 23]. According to [22], this instability requires the direction of the electric field to be the same as the electron density gradient in the region of formation of instabilities. To examine this possibility, we have shown the magnitudes of the x and y-components of $\vec{E} \cdot \nabla n_e$ in Figures 12(a) and 12(c), respectively. The sign of the x and y-components is also plotted in Figures 12(b) and 12(d), respectively. We observe that the sign of the dominant y-directed electric field is opposite to the sign of electron density gradient in the y-direction in the bottom region where there is a sheath above RF electrode. This is also the case for the x-component too except at the left and right corners. However, in these two corners, the magnitude of the product of the x-directed field and the x-directed density gradient is not significant as seen in Figure 12(a) for magnetic fields ranging from 30 G to 100 G. This violates the condition for the Simon-Hoh or modified Simon-Hoh instabilities (MSHI).

The disappearance of instabilities near the left bottom is possibly because the electrons are lost to the ground wall where the sheath is thin and electron cyclotron motion is interrupted by the walls. With the loss of a considerable number of electrons from the spoke-like structures to the ground wall, we expect the instabilities to disappear. This effect of reduction of spoke-like $\vec{E} \times \vec{B}$ instabilities at metal boundaries is discussed in [43].

In [44], Xu et al. investigated an m = 1 rotating spoke in a magnetically enhanced cathode arc discharge using 2D particle in cell (PIC) simulations and observed instabilities that begin in anode sheath (where quasi-neutrality is violated) that have a significant azimuthal and negligible radial component in wave number very similar to the instabilities seen in this paper. The wavelength observed for the instabilities in this paper are in the range of a few millimeters. For the 75 G case at 10 mTorr and 1 nF capacitor where the frequencies of instabilities are evaluated to be close to 2 MHz at x = 1.25 cm and y = 3.75 cm, the electron Larmor radius is estimated as 0.576 mm. This is neither in the long wavelength regime nor in the small wavelength regime. In the long wavelength regime, such instabilities are mentioned to be Simon-Hoh instabilities (SHI) in [44] which we have established are not the instabilities observed in this paper. In the small wavelength regime, such instabilities are mentioned to be either lower hybrid instabilities (LHI) or ion sound instabilities (ISI) in [44].

In [28], Lucken et al. describe instabilities in RF plasma similar to those seen in our paper. Lucken et al. investigated the instabilities using a 1D model of cross field plasma transport and validated using 2D PIC simulations. The 1D model in [28] describes that the sheath edge is almost always unstable if electrons are sufficiently magnetized, which is consistent with our observation for magnetic fields >= 75 Gauss at 10 mTorr pressure. The instabilities mentioned in [28] have wavelengths of the order of a few millimeters, begin at the sheath edge, and have a significant azimuthal wave number component, just like the instabilities in our paper. In [28], the criterion for instability is that the diamagnetic drift or fluid electron drift is higher than the ion sound speed. The instabilities in our paper rotate in the direction of diamagnetic drift, which is opposite to the direction of $\vec{E} \times \vec{B}$ drift. In addition, the diamagnetic drift velocity is significantly greater than the ion sound speed at 75 G and 100 G magnetic field where we observe the instabilities. The instability waves have a continuous spectrum that peaks at 3 MHz in [28]. The instability waves in our simulations have a continuous spectrum with a peak around 2 MHz, indicating another similarity.

We further analyze the 75 G, 1 nF cases at 10 mTorr and 20 mTorr pressure using the dispersion relation. The two-fluid linear dispersion equation with collisions included has the form [44, 45, 46, 47] given in the following equation:

$$k^2 \lambda_{De}^2 = \frac{\alpha k^2 c_s^2}{\omega^2} - \frac{\omega_d + k^2 \rho_e^2 (\omega - \omega_E + i\nu_{en})}{\omega - \omega_E + k^2 \rho_e^2 (\omega - \omega_E + i\nu_{en})} \tag{1}$$

In Equation (1), $k$ is the angular wave number and $\omega = \omega_r + i\,\omega_i$ is the angular frequency with real and imaginary parts predicting the instability wave frequency and the instability growth rate, respectively. The other terms in Equation (1) are defined in [44]. In Equation (1), it is assumed that the ions are non-magnetized. The instability begins in the bottom-right sheath region where quasi-neutrality condition is violated and hence, we have included the non-neutrality effect. The term on the left of Equation (1) represents the Debye-length effect coming from the Poisson equation whereas the terms on the right represent ion and electron responses.

| Plasma quantity | Value in PIC simulation (10 mTorr) | Value in PIC simulation (20 mTorr) |
|---|---|---|
| Electron density, $n_e$ | 4.98x10$^{15}$ m$^{-3}$ | 8.17x10$^{15}$ m$^{-3}$ |
| Electron temperature, $T_e$ | 3.26 eV | 2.54 eV |
| Ion density, $n_i$ | 5.06x10$^{15}$ m$^{-3}$ | 8.16x10$^{15}$ m$^{-3}$ |
| Electron density gradient length scale, $L_n$ | 11.25 mm | 10.36 mm |
| Electric field strength, $E_0$ | 255.54 V/m | 719.75 V/m |
| Electron-neutral collision frequency, $\nu_{en}$ | 35.233 MHz | 61.093 MHz |

**Table 1**: The plasma quantities extracted at the probe position in the top-left corner (x = 1.25 cm, y = 3.75 cm) from the PIC simulation of the 75 Gauss, 1 nF cases for 10 and 20 mTorr.

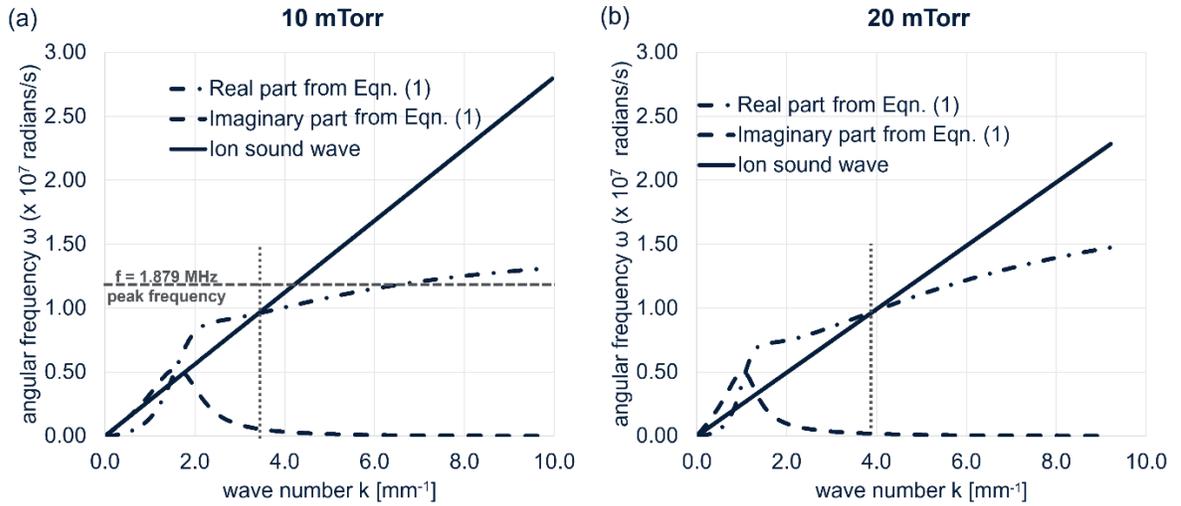

**Figure 13**: The real and imaginary parts of angular wave frequency corresponding to wave number obtained from Equation (1) for the 75 Gauss, 1 nF case at (a) 10 mTorr and (b) 20 mTorr pressures with plasma quantities extracted from the PIC simulation at probe position in the top-left corner (x = 1.25 cm, y = 3.75 cm). The wavenumber at which the ion sound speed equals wave velocity from Equation (1) is marked with dotted vertical lines in both plots (a) and (b) above. The peak frequency 1.879 MHz of the continuous spectrum obtained for the instabilities in 75 G, 10 mTorr, 1 nF case displayed in Figure 11(b) is marked on plot (a) above with a dashed horizontal line.

Table 1 contains the plasma quantities obtained from the PIC simulation for the 75 G, 1nF case at the probe position located at the left-top corner (x = 1.25 cm, y = 3.75 cm). These data after averaging over one RF cycle for 10 and 20 mTorr are located in the second and third columns, respectively. Using the quantities from Table 1, the dispersion relation ($\omega_r$ and $\omega_i$ vs $k$) in Equation (1) is plotted in Figure 13(a) and (b) for 10 and 20 mTorr pressures, respectively. $\omega_r$ from Equation (1) is compared to that of the ion sound speed at the same $k$. At the wavenumber $k$ where $\omega_r$ of the ion sound wave is small compared to $\omega_r$ from Equation (1), the ion sound speed is less than the wave phase velocity given by Equation (1). We expect to see instabilities for which the growth rate of instabilities, $\omega_i$, given by Equation (1), as evident in Figure 13(a) where the threshold wavenumber for instability is marked with a dotted vertical line.

If we compare the 20 mTorr case with the 10 mTorr case, the instability growth rate $\omega_i$ is reduced due to the effect of collisions, as shown in Figure 10(e) and 10(f). The frequency calculated from analytical dispersion relation Equation (1) is also close to the peak frequency in the wave spectrum obtained at in the PIC simulations shown in Figure 11(b). This peak frequency is marked on the plot in Figure 13(a) using a horizontal dashed line. This indicates that the region of the continuous spectrum around the peak frequency coincides with the angular frequency where ion sound speed equals wave velocity from Equation (1) showing additional proof of why instabilities are seen in the 75 G, 10 mTorr, 1 nF case. Note that established turbulence is governed by nonlinear processes and cannot be described by the linear dispersion relation, so this is just a guidance given by the heuristic argument. The 20 mTorr case shows some positive growth ($\omega_r$) according to the linear dispersion relation as seen in Figure 11(b) but the instabilities are not clearly visible in the simulation results as seen in figure 8(e) which indicates that the turbulence is highly nonlinear and requires further investigation into the dispersion relation without any assumption of linearity.

The configuration of the electric and magnetic fields in our paper is similar to the configuration used in the MISTRAL device described in Refs. [48, 49, 50], even though the MISTRAL device is not a CCP. The axial magnetic field in the long cylindrical plasma source in the MISTRAL device enabled $\vec{E} \times \vec{B}$ drift and the rotating instabilities look similar to those described in this article. The Penning discharges, see e.g., Ref. [51], also enables similar $\vec{E} \times \vec{B}$ drift and could lead to instabilities as seen in our simulations. Both these experimental setups are good candidates to experimentally validate instabilities reported in this article.

## 5. Conclusion

In this paper, 2D PIC simulations are performed in an asymmetric CCP discharge with Ar gas at 10 – 100 mTorr gas pressure. A 40 MHz RF power source is used to ignite plasma with a uniform magnetic field perpendicular to the plane of the 2D CCP discharge.

At low magnetic fields (up to 50 Gauss at 10 mTorr), a shift in plasma profile from the center towards one corner of the biased electrode is observed. The plasma profile shifts more with increasing magnetic field. The addition of an external capacitor resulted in increased sheath potential and reduced electron density for magnetic field strengths up to 50 G. This was termed as the first regime of plasma behavior with increasing magnetic field strength.

At higher magnetic fields (75 Gauss for 10 mTorr), we observe instabilities resembling spoke-like self-organized structures within the sheath that rotate around the chamber in anti-clockwise or $-\vec{E} \times \vec{B}$ direction. This was called the second regime of plasma behavior with increasing magnetic field strength.

At even higher magnetic fields (100 Gauss at 10 mTorr), the coherent instabilities continued to exist within the sheath, but new unstable structures of different shapes and sizes were formed in the bulk plasma region moving in an incoherent fashion. This was termed as the third regime of plasma behavior with increasing magnetic field strength.

Addition of external capacitor led to an increase in electron density in the second and third regimes, as explained by a larger source of electrons in the results with the capacitor. Addition of capacitor did not change the thresholds between regimes of plasma behavior. Reducing area ratio between the RF powered to ground electrodes led to shrinking of plasma source volume due to increased asymmetry but did not change thresholds between regimes of plasma behavior.

Increase in pressure, however, led to increase in electron-neutral collisions and damping of the instabilities. Therefore, increasing pressure led to increase in threshold values of the magnetic field between the different regimes of plasma behavior.

The instabilities within the sheath rotate at frequencies close to 2 MHz. The instabilities are discovered to be neither electron cyclotron drift instabilities nor modified Simon-Hoh instabilities. However, the properties of the instabilities observed in this paper do resemble previously studied plasma instabilities reported in Ref. [28]. A 2D PIC model with both uniform magnetic and RF inductive electric fields was used in Ref. [22] instead of nonuniform capacitive RF field studied here. Because this paper describes the first self-consistent asymmetric CCP plasma model with non-uniform electric field that shows evidence of instabilities studied in [28], this paper can pave the way for experimental validation of such instabilities. The MISTRAL device in [48, 49, 50] and the cold cathode ion source in [51] are good candidates for the validation study if required adjustments (additional RF electrodes are added) are made to the experimental setup. Additionally, this paper provides threshold values of magnetic field for the appearance of instabilities in RF plasma so that such process space can be avoided in semiconductor applications where magnetized plasma at low pressure is employed to achieve spatially uniform high density thin film deposition / high energy etch on the wafer.

## 6. Acknowledgement

We would like to thank Peng Tian for assistance with EDIPIC simulations and Sarvesh Sharma for fruitful discussions.

The authors of this paper have no potential conflict of interest to disclose.

This work at Princeton Plasma Physics Laboratory (PPPL) was supported by US Department of Energy CRADA agreement between Applied Material Inc. and PPPL.

**Appendix**

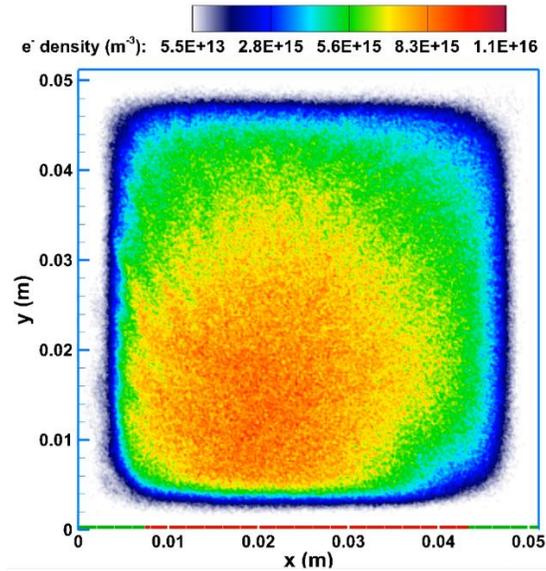

*Figure 14: The screen shot of the video (**video will be made available by authors on reasonable request**) showing rotation of spoke-like instabilities emanating from bulk plasma center to the process region boundaries for PIC simulation with applied magnetic field of 75 G in a chamber with Argon gas at 10 mTorr driven by 200 V amplitude RF voltage at 40 MHz through no blocking capacitance and RF powered to ground electrode area ratio of 0.25*

The videos of electron density are included in this appendix for several conditions. The electron density is plotted every 80 RF cycles in these videos. $n_e$ is shown in Figure 14 (screenshot for video) for 10 mTorr pressure, no blocking capacitor, 200 V peak voltage, RF powered to ground electrode area ratio of 0.25, and magnetic field of 75 G. Spoke-like instabilities emanating from bulk plasma at the center to the edge of the process region can be seen in this video. The spokes are rotating around the bulk plasma in the anti-clockwise ($-\vec{E} \times \vec{B}$) direction.

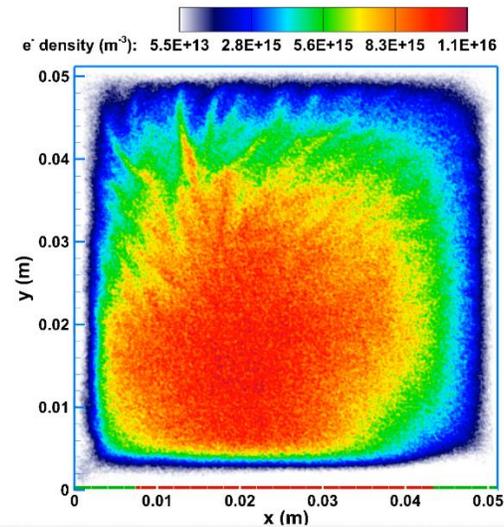

*Figure 15: The screen shot of the video (**video will be made available by authors on reasonable request**) showing rotation of spoke-like instabilities emanating from bulk plasma center to the process region boundaries for PIC*

*simulation with applied magnetic field of 75 G in a chamber with Argon gas at 10 mTorr driven by 200 V amplitude RF voltage at 40 MHz through a 1 nF blocking capacitance and RF powered to ground electrode area ratio of 0.25*

To illustrate the effect of the blocking capacitor, $n_e$ is shown in Figure 15 (screenshot for video) for 10 mTorr pressure, 1 nF blocking capacitor, 200 V amplitude voltage, RF powered to ground electrode area ratio of 0.25, and magnetic field of 75 G. Spoke-like instabilities emanating from bulk plasma at the center to the edge of the process region can be seen in this video. The spokes are rotating in the anti-clockwise ($-\vec{E} \times \vec{B}$) direction as in the case without the capacitor except that the plasma has drifted more to the left and bottom compared to the case without capacitor.

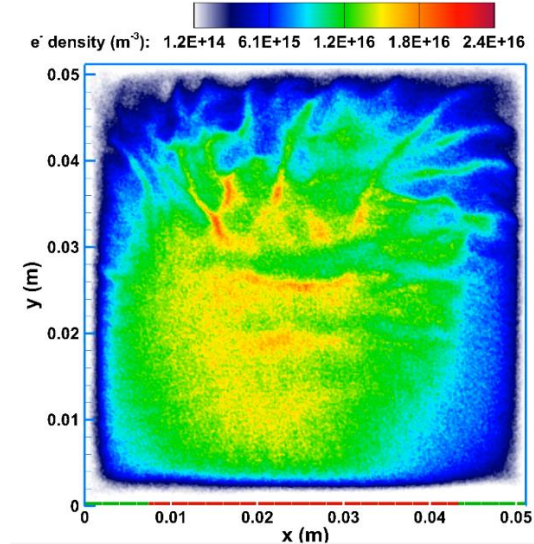

*Figure 16: The screen shot of the video (**video will be made available by authors on reasonable request**) showing rotation of spoke-like instabilities emanating from bulk plasma center to the process region boundaries for PIC simulation with applied magnetic field of 100 G in a chamber with Argon gas at 10 mTorr driven by 200 V amplitude RF voltage at 40 MHz through a 1 nF blocking capacitance and RF powered to ground electrode area ratio of 0.25*

To examine the effect of increased magnetic field, video of $n_e$ is shown in Figure 16 (screenshot for video) for 10 mTorr pressure, 1 nF capacitor, 200 V amplitude voltage, RF powered to ground electrode area ratio of 0.25, and magnetic field of 100 G. Instabilities of different shapes at different locations within the process region can be seen in this video. Both linear and rotational types of movement within plasma process region can be observed.

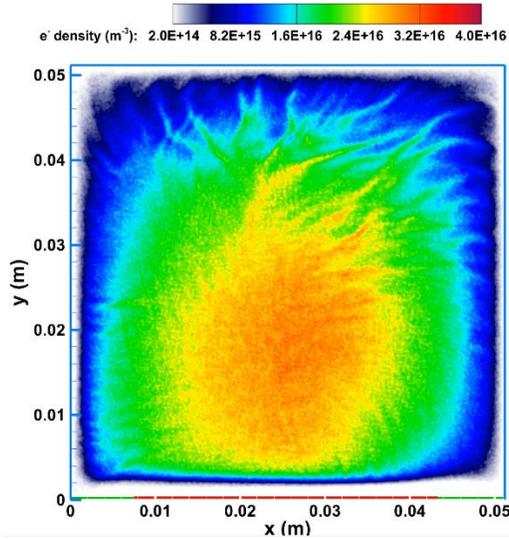

*Figure 17: The screen shot of the video (**video will be made available by authors on reasonable request**) showing rotation of spoke-like instabilities emanating from bulk plasma center to the process region boundaries for PIC simulation with applied magnetic field of 100 G in a chamber with Argon gas at 20 mTorr driven by 200 V amplitude RF voltage at 40 MHz through a 1 nF blocking capacitance and RF powered to ground electrode area ratio of 0.25*

The effect of increasing gas pressure can be viewed in the video in Figure 17 (screenshot for video). Conditions in this video are 20 mTorr pressure, 1 nF capacitor, 200 V amplitude voltage, RF powered to ground electrode area ratio of 0.25, and magnetic field of 100 G. Spoke-like instabilities emanating from bulk plasma at the center to the edge of the process region can be seen in this video. These spoke-shaped structures are rotating around the bulk plasma in the anti-clockwise direction, very similar to the 75 G case without capacitor above, except that the plasma seems to be more centered in this case without a big noticeable drift.

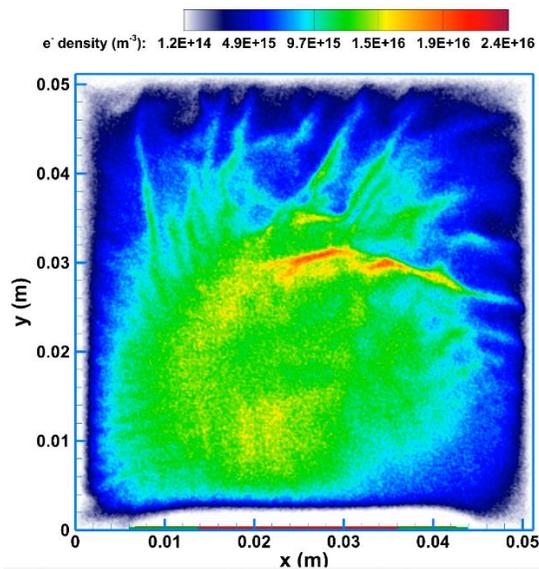

*Figure 18: The screen shot of the video (**video will be made available by authors on reasonable request**) showing rotation of spoke-like instabilities emanating from bulk plasma center to the process region boundaries for PIC*

*simulation with applied magnetic field of 100 G in a chamber with Argon gas at 10 mTorr driven by 200 V amplitude RF voltage at 40 MHz through a 1 nF blocking capacitance and RF powered to ground electrode area ratio of 0.15*

Video of $n_e$ at the smaller RF powered to ground area ratio is included in Figure 18 (screenshot for video). The conditions are 10 mTorr pressure, 1 nF capacitor, 200 V amplitude voltage, RF powered to ground electrode area ratio of 0.15, and magnetic field of 100 G. Instabilities of different shapes at different locations within the bulk plasma region can be seen in this video. These spoke-like structures have both linear and rotational types of movement within the bulk plasma region and in the thick sheath region respectively. Also, the plasma profile at the bottom where the RF electrode is narrower seems to be different compared to the 100 G case with electrode area ratio of 0.25.

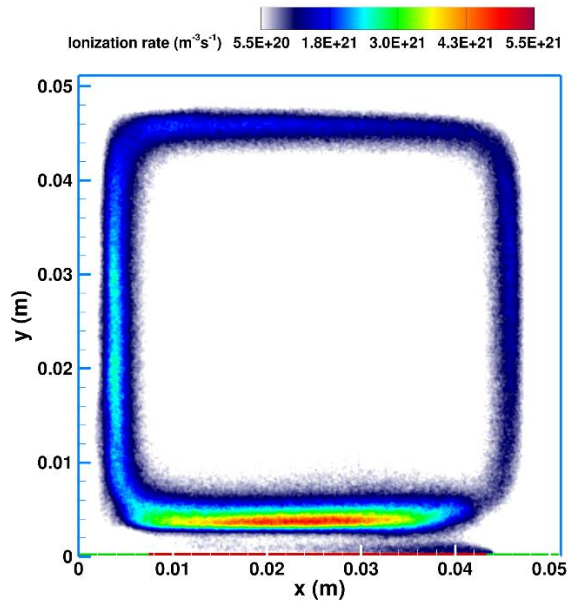

*Figure 19: Time-averaged ionization rate whose time variation is shown in video (**video will be made available by authors on reasonable request**) where ionization rate is peaking above bottom RF electrode for PIC simulation with applied magnetic field of 75 G in a chamber with Argon gas at 10 mTorr driven by 200 V amplitude RF voltage at 40 MHz with no blocking capacitance and RF powered to ground electrode area ratio of 0.25.*

Video of ionization rate for 75 Gauss magnetic field, 10 mTorr pressure and no blocking capacitor is included in Figure 19 (figure is time-averaged result over 80 RF cycles). The RF powered to ground electrode area ratio of 0.25. Ionization rate peaks above RF powered electrode where the sheath thickness is large comparatively due to asymmetric electrode areas. The ionization rate is also slightly higher next to the left boundary compared to the top and right boundary as there is a leftward $\vec{E} \times \vec{B}$ drift of bulk plasma.

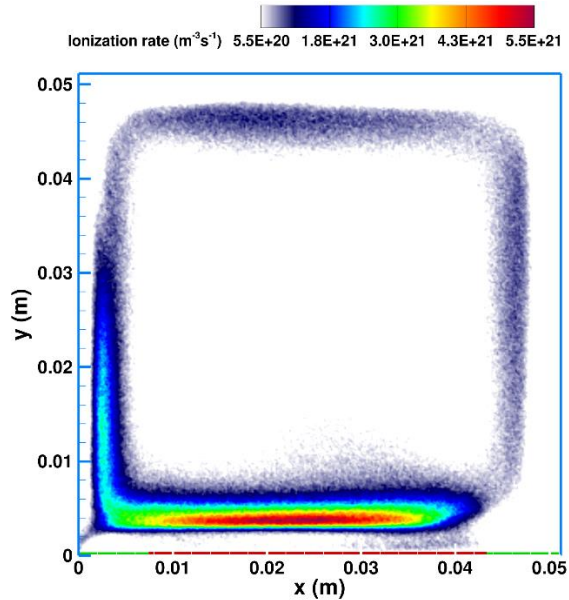

*Figure 20: Time-averaged ionization rate whose time variation is shown in video (**video will be made available by authors on reasonable request**) where ionization rate is peaking above bottom RF electrode for PIC simulation with applied magnetic field of 75 G in a chamber with Argon gas at 10 mTorr driven by 200 V amplitude RF voltage at 40 MHz through a 1 nF blocking capacitance and RF powered to ground electrode area ratio of 0.25.*

Video of ionization rate for 75 Gauss magnetic field, 10 mTorr pressure and 1 nF blocking capacitor is included in Figure 20 (figure is time-averaged result over 80 RF cycles). The RF powered to ground electrode area ratio of 0.25. Ionization rate peaks above RF powered electrode where the sheath thickness is large comparatively due to asymmetric electrode areas. The peak ionization continues leftward to the region above the bottom left vacuum gap. With a blocking capacitor, the RF powered electrode develops a negative voltage and this leads to electric field from ground to RF powered electrode at the left bottom boundary which is not present in the case without blocking capacitor. This extra field helps in extra ionization above the bottom left boundary in the 1 nF capacitor case compared to no capacitor case. This explains the higher overall electron density seen in the 75 Gauss, 10 mTorr case with 1 nF capacitor compared to the 75 Gauss, 10 mTorr case without capacitor. This is probably the reason why 100 Gauss, 10 mTorr case with 1 nF capacitor has a higher overall electron density compared to the 100 Gauss, 10 mTorr case with no capacitor. The ionization rate is also slightly higher next to the left boundary compared to the top and right boundary as there is a leftward E x B drift of bulk plasma.